\documentclass{article}

\usepackage{PRIMEarxiv}

\usepackage[utf8]{inputenc} 
\usepackage[T1]{fontenc}    
\usepackage{hyperref}       
\usepackage{url}            
\usepackage{booktabs}       
\usepackage{amsfonts}       
\usepackage{nicefrac}       
\usepackage{microtype}      
\usepackage{lipsum}
\usepackage{fancyhdr}       
\usepackage{graphicx}       
\graphicspath{{media/}}     
\usepackage[inline]{enumitem}
\usepackage[disable]{todonotes}

\title{DevOps and Microservices in Scientific System development: experience on a multi-year industry research project
\thanks{\textit{\underline{Citation}}: 
\textbf{Maximillien de Bayser, Vinicius Segura, Leonardo Guerreiro Azevedo, Leonardo P. Tizzei, Raphael Melo Thiago, Elton Soares, and Renato Cerqueira. Devops and Microservices in Scientific System Development: Experience on a Multi-year Industry Research Project. In: The 37th ACM/SIGAPP Symposium on Applied Computing, DOI:10.1145/3477314.3507317.}} 
}

\author{
  Maximillien de Bayser, Vin\'{i}cius Segura, Leonardo G. Azevedo \\
  IBM Research \\
  Rio de Janeiro, Brazil\\
  \texttt{\{mbayser, vboas, lga\}@br.ibm.com} \\
   \And
  Leonardo P. Tizzei \\
  IBM Research \\
  São Paulo, Brazil\\
  \texttt{ltizzei@br.ibm.com} \\
   \AND
  Raphael Thiago, Elton Soares, Renato Cerqueira \\
  IBM Research \\
  Rio de Janeiro, Brazil \\
  \texttt{\{raphaelt, rcerq\}@br.ibm.com, eltons@ibm.com} \\
}

\begin{document}
\maketitle

\begin{abstract}
There is a gap in scientific information systems development concerning modern software engineering and scientific computing.
Historically, software engineering methodologies have been perceived as an unwanted accidental complexity to computational scientists in their scientific systems development. 
More recent trends, like the end of Moore's law and the subsequent diversification of hardware platforms, 
combined with the increasing multidisciplinarity of science itself have exacerbated the problem because self-taught ``end user developers'' are not familiar with the disciplines needed to tackle this increased complexity. 
On a more positive note, agile programming methods have approached software development practices to the way scientific software is produced. 
In this work, we present the experience of a multi-year industry research project where agile methods, microservices and DevOps were applied.
Our goal is to validate the hypothesis that the use of microservices would allow computational scientists to work in the more minimalistic prototype-oriented way that they prefer while the software engineering team would handle the integration.
Hence, scientific multidisciplinary systems would gain in a twofold way: 
\begin{enumerate*}[label=(\roman*)]
    \item Subject Matter Experts (SME) use their preferable tools to develop the specific scientific part of the system; 
    \item software engineers provide the high quality software code for the system delivery.
\end{enumerate*}
\end{abstract}


\section{Introduction}
\label{sec:intro}

\todo[inline]{text commented due to space restrictions}
``Simple things should be simple, complex things should be possible.''
This maxim by Alan Kay is considered a guiding principle by designers of programming languages, libraries,  applications, and any other artefacts that might be considered to be tools. But it begs the questions: Simple or possible for whom? What is
simplicity or complexity? Specifically in the domain of information systems, who are the tool designers and who are the
users? 

In this work, we consider ourselves, the software engineers, as tool designers and computational scientists as
``professional end user developers''~\cite{4351335}. 
Contrary to our intention as software engineers, extensive documentation in literature shows that computational scientists, who regard software development as a means to an end, perceive the methodologies of software engineering
as accidental complexities that stand in the way of science~\cite{Johanson,Storer:2017:BCS}.

At the risk of sounding unpoetic, perhaps we should rephrase the maxim as ``Simple things should be easy, complex things should
be not too hard.'' to separate the complexity and difficulty dimensions and allow a more profound analysis. 
Difficulty is a human dimension related to the amount of effort required to overcome perceived obstacles. 
Segal~\cite{4351335} reports that scientists in general have no difficulty in learning programming languages by themselves, 
especially in languages that have a large body of documentation and a previous history in scientific computing. 
However, evidence such as the scandal about errors in climatic change prediction software by the University of East Anglia~\cite{anglia} suggest that scientists are struggling with software quality in large projects. 
In contrast, software engineers are often unable to implement scientific software following specifications by scientists~\cite{4222616}.

Complexity is a cause of perceived difficulty in intellectual activities. To avoid a deeper philosophical discussion about complexity,
we will borrow the following definition:  ``Complexity characterises the behaviour of a system or model whose components interact in multiple
ways and follow local rules, meaning there is no reasonable higher instruction to define the various possible interactions.''~\cite{wiki:Complexity}.

This definition is in line with definitions of complexity in software engineering like the cyclomatic complexity~\cite{1702388} or Halstead's metrics~\cite{halstead}.
Complexity can be divided into incidental and accidental.

Johanson and Hasselbring~\cite{Johanson} identify two trends that have increased the complexity of scientific computing in recent years:
\begin{enumerate}
 \item \textbf{Increase in incidental complexity:} As science itself is evolving and requiring more interdisciplinarity, so does scientific software. It is now common
 to couple physical models developed by different research disciplines. An example is the simulation of fluid flow
 in a petroleum reservoir coupled with the models of chemical reactions. While scientist in the past might have been able to
 ``code'' on their own, now they must integrate with the work of others.
 \item \textbf{Increase in accidental complexity:} It is the nature of scientific computing that any advance in hardware is taken
 advantage of to run more accurate models. During the years of Moore's law, scientists could rely on increasing CPU speeds for more
 computing power. With the end of this trend the adoption of heterogeneous architectures with multicore processors, distributed computing, and specialised
 co-processors like GPUs have become a necessity. Scientists however are ill-prepared to deal with this new reality.
\end{enumerate}

In the past, upfront design software engineering practices 
were at contrast with scientific
development practices. 
A theory about specific phenomena is developed by a process of formulating hypotheses and running experiments
to try to invalidate those hypotheses. 
Whenever a contradiction is found, the model must be adjusted and subjected to further experiments,
resulting in an iterative process. 
Since scientists view their code as a representation of their evolving theory, it is clear that no complete specification of the software can be written upfront. 

However, in recent years, software engineering has adopted more iterative approaches like Agile, to mitigate the risk that a software is found to be at odds with real-world requirements after years in development. 
Several studies applying agile methodologies to scientific computing have been documented~\cite{Ackroyd,Kleb}. 
Practices such as continuous integration (\texttt{CI}) foster an experimental approach to development where small incremental changes are tested under real conditions with staging, A/B testing, and extensive monitoring \cite{savor2016continuous}.
Moreover, the dynamism of development and stability required for operation generates conflicts among areas, teams and goals, demanding changes in people mentality and reasoning about the impacts in organizational work culture~\cite{gimenez2020devops}.
Therefore, tools and methods are required to meet the needs of scientists and engineers in scientific system development.

In this work, we explore the development of scientific system applying recent software engineering trends such as 
microservices~\cite{fowler2014microservices} supported by \texttt{DevOps}~\cite{huttermann:2012:devops} practices and as-a-service delivery. 
We support our findings on our experience of applying these trends from the start to a large applied research project involving a multidisciplinary team of more than 30 people over 3 years. 
In this project, in the domain of petroleum geology and geophysics, several microservices were developed by small teams composed of subject matter experts (SMEs) and developers. 
\texttt{DevOps}, \texttt{CI} and user interface were handled by our software engineering staff. 
\todo[inline]{We nicknamed the overall methodology ``ResearchOps''~\cite{DeBayser:2015:ResearchOps}}

It is a known property of microservices that they promote a different trade-off in the complexity of the overall
system: the individual components become simpler, and the integration of parts becomes more complex~\cite{newman2015building,zimmermann2016microservices}.
The hypothesis that we wanted to test in this project was that this trade-off would allow us to relieve SMEs to some degree from the burden of
dealing with the complexity of software and shift it to the software engineering staff. 
The SME would use the programming language that is most familiar to scientists in her domain of expertise and her work would be integrated into the larger application using standard interfaces. 
In other words, we expected that microservices would allow us to re-distribute the complexity so as to minimise the overall difficulty.

This paper is organised as follows. 
Section~\ref{sec:the-project-development} presents how we have implemented microservices, \texttt{DevOps}, and \texttt{CI}.
Section~\ref{sec:study} presents the quantitative and qualitative metrics that we have gathered about the prototype and the teams that implemented it.
Section~\ref{sec:related} discusses some related works.
Finally, section~\ref{sec:conclusion} presents our conclusions.

\section{Incentive structures in scientific computing}
\label{sec:incentives}

According to Storer \cite{Storer:2017:BCS} and Johanson \cite{Johanson}, the value of software for scientists lies solely in the amount of publishable
scientific results that it generates per calendar time. 
Therefore shortcuts are taken to reduce the time to scientific results, be it by optimisation that reduce running time but also reduce portability or reuse, 
or by not investing time to modularise the code. 
The grant-based funding in academia c`ontributes to this because it casts uncertainty about the continuation of lines of research in the future. 
It also makes it hard to justify the employment of software engineers to help the scientists although it has repeatedly been demonstrated that it improves outcomes \cite{5337641}.

As in other industry research institutes, we do have a permanent software engineering staff that supports and orients the development of projects. 
We also have a long term research strategy and, therefore, value the re-usability of the software artifacts that are produced. 

However, as is the case in academia, the performance of our scientists and the ones in the sponsoring organisations is still measured by metrics like the number of patents and papers produced.
This means that new features tend to be valued more than quality and performance improvements of existing code.
And as in academia, researchers are already very busy keeping up with their fields and tend to resist learning more advanced software engineering concepts.

As a consequence, the engineering team has had difficulties in controlling the proliferation of \emph{ad hoc} solutions. 
As a concrete example, to reduce the size of arrays stored in a \texttt{MongoDB} \cite{MongoDB} database, 
the native representation of arrays was exchanged by a binary blob of serialised (\emph{pickled}) \texttt{Python} arrays. 
Although it reduced storage requirements in half, it made the database effectively unreadable for other programming languages, 
including other versions of \texttt{Python}. 
This is a good example of something that looks like an easy solution, but hides a potentially large ``technical debt''. 
We have had other instances of this problem such as in the serialisation of a machine learning model for image classification.

As software engineers, we should improve efforts to explain to stakeholders how the fast addition of features correlates negatively with code quality and therefore stability, performance, and reliability of software artefacts can suffer.
Because without this awareness, end user developers might miss the chance to seek advice before committing to a bad design.
\section{The project development}
\label{sec:the-project-development}
This section presents how the project was developed.

\subsection{The Project}
\label{sec:project}
The project that is the object of this study was a multi-year joint research agreement between our research institute and an Oil\&Gas company.
On the client side, we collaborated with a team of geologists and geophysicists. 
On our side, we had experts in fields such as optimisation, computer vision, machine learning and, visualisation. 
Our experts had the support of interns and our software engineering staff. 

The project aimed at integrating research in several different aspects of petroleum geophysics. 
Each of these aspects, that we called \emph{workstreams}, had an assigned team which worked quite independently, therefore warranting the use of microservices.

The client SMEs did not directly participate in software development, but oriented the research and evaluated results. 
The system was delivered to them as-a-service prototype with a web-based user interface. 
We effectively employed \texttt{CI} to allow quick iterations.

\subsection{Architecture}
\label{sec:architecture}

A microservice architecture comprises independently developed services that work together to achieve a specific purpose~\cite{newman2015building}. 

Microservice architecture has become the established approach for a huge set of modern applications, as for the well-known companies like Netflix and Amazon.
It is also an approach that should be considered for a new software lifecycle of a legacy enterprise application or its redevelopment.
Microservices can meet the new demands and responds as flexible as possible to changes in the future.
It is able to handle this in the technical and the process level serving as the basis for future concepts~\cite{stricker2018microservices}.

The general architectural pattern for our microservices is to expose a RESTful \cite{Fielding:2000:ASD} HTTP interface.
The advantages of using HTTP as a communication protocol are:

\begin{enumerate}
 \item \textbf{Language Independence:} all popular programming languages have libraries for HTTP clients and servers;
 \item \textbf{Human readable:} HTTP requests and responses in text format allow developer inspection without specialised tools;
 \item \textbf{Tooling:} there is a large amount of tooling available for HTTP, from web browsers to server software, load balancers and proxies.
\end{enumerate}

There also are several downsides:

\begin{enumerate}
 \item \textbf{Text-based:} when transmitting data that is not natively represented as text, client and server spend CPU cycles converting data to text. Caching
 can help to mitigate this overhead;
 
 \item \textbf{Latency:} every new request has to go through the three-way HTTP handshake and, slow start. With HTTPS it gets even worse. This
 can be mitigated using \emph{Keep-Alive}, but only when the cost can be amortised over several requests;
 
 \item \textbf{Bandwidth:} text is much more wasteful in bandwidth than compact binary protocols, if not using compression;
 
 \item \textbf{Interaction pattern:} the only interaction pattern is request-response, which limits more complex interactions. 
\end{enumerate}

All microservices are written in \texttt{Python}, \texttt{Javascript} or \texttt{Java}. Horizontal concerns like request authorisation, logging, encryption (Transport Layer Encryption\footnote{https://tools.ietf.org/html/rfc8446}),
load-balancing and caching are all implemented by putting a \texttt{nginx} server as reverse proxy in front of these microservices. This approach, besides the advantage
of uniform request logging and centralised security, is one of the possible complexity trade-offs that allow us to simplify the implementation of microservices and
transfer the burden to the integration team.

Since we do not operate at a large scale and all microservices are part of the same application, all microservices share
the same replicated \texttt{nginx} servers, but we could equally well use the \emph{sidecar pattern}~\cite{Burns:2018:DDS} and deploy one \texttt{nginx} for each microservice. This would
lead to a greater configuration complexity but would give us fine-grained control over individual services.

Authentication is handled by one API endpoint of a specific microservice that returns \texttt{JWT} tokens \cite{JWT}. A script written in \texttt{Lua} and run by \texttt{nginx}
then enforces that all other requests include this token in a \texttt{HTTP} header.

In \texttt{Python}, on the developers machine we use the \texttt{Flask} \texttt{HTTP} server library, but in production we switch to the \texttt{uwsgi} server 
software that provides concurrency by starting multiple \texttt{Python} processes and also supports the \texttt{WSGI} binary protocol to communicate more efficiently
with the \texttt{nginx} reverse proxy. This is yet another example of an effective complexity trade-off.

\subsubsection{API Gateways}

\begin{figure}[!htb]
\centering
 \includegraphics[width=0.4\textwidth]{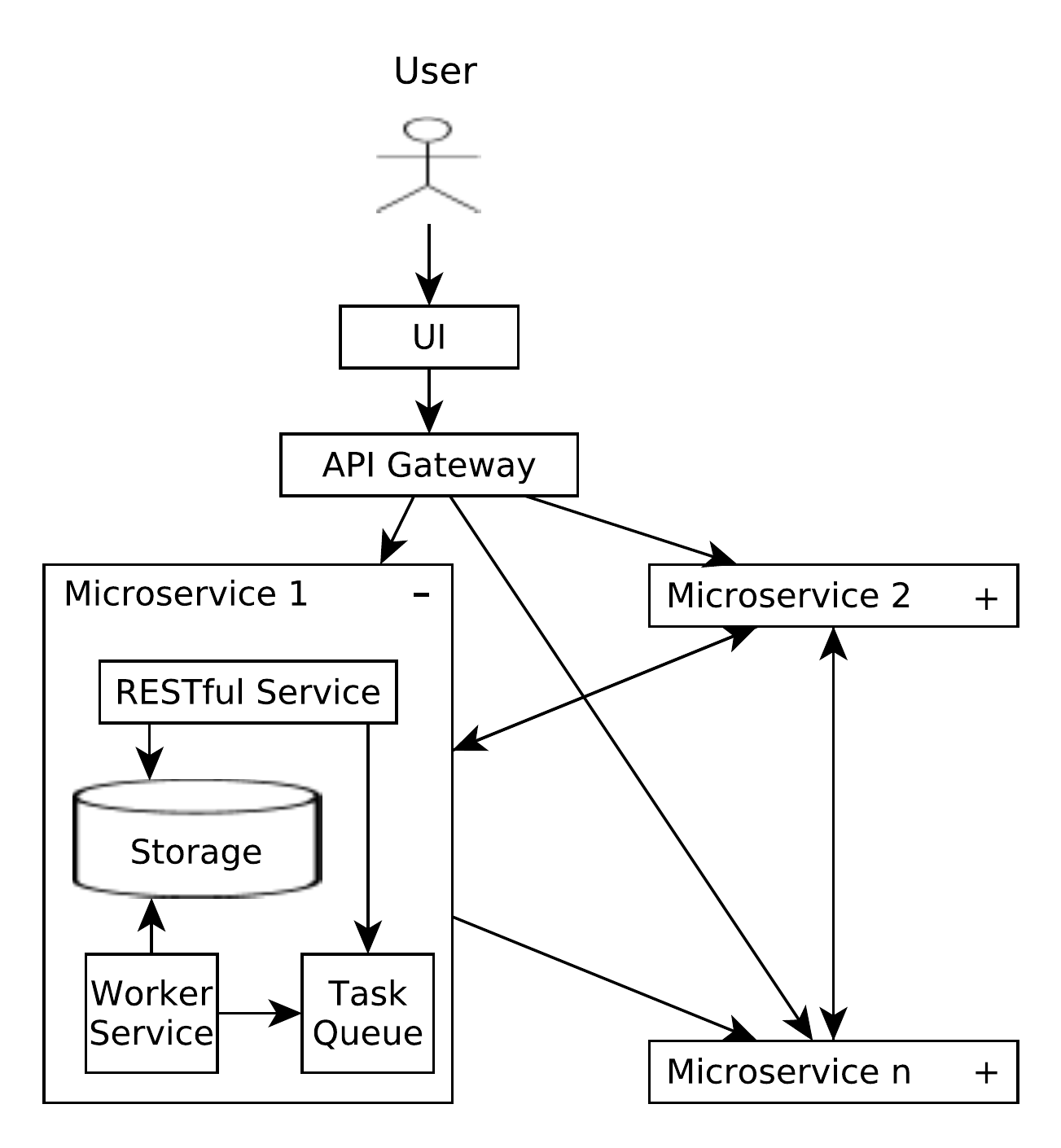}
 \caption{Microservices in detail}
 \label{fig:ms_diagram}
\end{figure}

In several instances, the \texttt{API} exposed by individual microservices is too fine-grained for the \texttt{UI} to consume. 
Therefore we have some services that act as \emph{API gateways} \cite{newman2015building}, as shown in Figure \ref{fig:ms_diagram}. 
These compose operations into larger ones that are called by the \texttt{UI}. 
This also shields the \texttt{UI} from \texttt{API} changes in the ``internal'' \texttt{APIs}, which is good because the team that develops it is not typically involved in the back-end services, while the other teams share the responsibility for the gateway service.

\subsubsection{Long running tasks}

Some microservices in the project must execute numerical algorithms on large seismic data. 
In production we have seen operations that took up to 6 days.
Clearly, the \texttt{HTTP} protocol was never meant to handle responses with such a large delay. 
The pattern we use to solve this is to return immediately a token that identifies the operation and start the operation in a background process. 
The token can then be used to query the state of the task and to obtain results when ready.

To implement this pattern, we use an encapsulated task queue in each of the microservices that need it, as shown in Figure \ref{fig:ms_diagram}. 
More specifically, we use the \texttt{celery} library for \texttt{Python} that handles the submission of tasks by master processes to worker processes over a queue server like \texttt{RabbitMQ}. 
Although each task queue is logically encapsulated, for deployment we use a single \texttt{RabbitMQ} server to host all task queues.

The combination of \texttt{RESTful} microservices with asynchronous background tasks has led to significant challenges to maintain encapsulation when the execution
of a task must be composed with calls to other microservices. 
The typical pattern that causes this difficulty is when a new request comes to the \texttt{API Gateway},
which in turn starts a long running task in service \texttt{A} and has to wait for it to end to compose the request with a book keeping operation in service \texttt{B}.
The \emph{ad hoc} solution we have thus far employed is to start so-called \emph{monitoring tasks} that wait for the end of background tasks and then finish whatever needs to be done. 

However, this approach has proved difficult to maintain because of fault-tolerance issues. 
A simple solution would be for \texttt{A} to call \texttt{B} directly at the end of the task, but this would break encapsulation and add circular dependencies between services, but it could be made to work adding the proverbial level of indirection through call-backs. 
We are still studying alternatives.

\subsubsection{Data processing pipeline}

In the last year of the project a data processing pipeline was added to the project. 
We mention it because it follows a different architectural pattern than
the other services. 
Pieces of data, submitted individually or in batches are submitted to a queue where they are picked up by services which in turn
extract and transform pieces of data and submit their results to other queues and so on. 
Some services just take the data off a queue and store it in a database, thus ending the pipeline.

In this pipeline, the only \texttt{HTTP} service is the submission service. 
From then on all communication between parts is done over \texttt{Kafka} \emph{topics}, which are a hybrid of queues and \emph{publish-subscribe}. 
The advantage of \texttt{Kafka} in this case is that workers can be scaled out and that we can re-play events in any depth of the pipeline due to the persistence of messages.

\subsection{Infrastructure}
\label{sec:infra}
While in the past the primary mode of delivery of our research group had been to install our software artefacts
directly on the user's workstation, this is the first large project that we have provided entirely as a service.
As a consequence we had to allocate and maintain an infrastructure and develop a deployment strategy.
While the deployment process is the subject of Section \ref{sec:devops}, this section discusses the infrastructure tools used.

\subsubsection{docker}

The project started in 2015, a time when container technologies started to gain more adoption in the industry~\cite{datadog}.
So when the first microservices were assembled from the proof-of-concept scripts that had been developed we started
to package them in containers.

The adoption of container technology, and \texttt{docker} \cite{DOCKER} in particular has brought us many benefits. 
It provides a standard medium to deploy software. 
All the \texttt{Dockerfiles} have been written by our software engineering team and we have written them in ways that allows the software to be configured by changing environment variables or overriding configuration files at run time. 
But we had to work closely with the microservices teams to remove any hard-coded locations of files and services.

However, there have been some difficulties as well. Some characteristics of the container model have been hard to grasp
for researchers without a computing background. In particular we could cite the ephemeral nature of containers. In several
cases we found that developers were writing to hard-coded locations in the container file system instead of notifying the
SE team that extra \emph{volumes} would have to be allocated and attached to running containers.
In general the whole lifecycle of containers, from the \texttt{docker build}, to the registry, the \texttt{docker} \emph{graph},
and, finally, the instantiation as a container contains several pitfalls for non-experts.

Operationally, the lack of maturity of docker has caused us problems on several occasions. It happened that the on-disk representation
of the \texttt{docker} \emph{graph} had changed after an upgrade and we had to erase everything to start again. Another problem
were deadlocks in the \texttt{docker} \emph{daemon} that forced us to restart everything.

According to literature surveys \cite{Storer:2017:BCS}, small changes in libraries or even compilers used to build scientific code are often
responsible for changes in the results, and therefore increase the difficulty to independently reproduce scientific results
that rely on software. 
\emph{Virtual machine} images built automatically in accordance with the
\emph{infrastructure as code} \cite{6265084} methodology could be a solution for this problem. 
Studies also found that \texttt{docker} images perform this role equally well, if not better due to the smaller size and overhead compared to \texttt{VM} \cite{Boettiger:2015:IDR:2723872.2723882}.
In addition, since there is no virtualisation involved, \texttt{docker} containers have direct access to hardware such as \texttt{GPUs}
or \texttt{Infiniband} when running on \emph{bare metal} server.

\subsubsection{docker-compose}

For several months we used a tool called \texttt{docker-compose} to orchestrate our containers. It was very convenient to start
or stop the entire prototype with a single command. And since we were not effectively using container tags at the time, we thought
it was convenient that the tool checked automatically if there was a new version of the image in the registry.

Another useful feature is that it can allocate a private virtual network on the host machine so that all containers run inside a single
segregated addressing space. This was especially convenient to us when we started to run several environments at the same time, as explained in
Section \ref{sec:devops}.

However, as the prototype grew too large for a single server we had to start looking for alternatives.

\subsubsection{kubernetes}

\texttt{Kubernetes} \cite{kubernetes, kubernetes2} was the container orchestration we chose to scale the project from a single workstation to a small server farm.
\texttt{Docker Swarm} at the time was too restrictive in features and \texttt{Mesos Marathon} was perceived as more complex on the basis of
being more than a container orchestrator.

Despite the steep learning curve at the beginning, our experience with \texttt{Kubernetes} has been a positive one. Multi-container
applications are created by submitting declarative specifications to a centralised \texttt{etcd} database. Several different kinds of
\emph{controllers} continuously monitor events to maintain an overall cluster \emph{homeostasis} by reacting to different
kinds of events. One kind of event is the submission of new specifications, other is changes in infrastructure
like the introduction or removal of computing nodes or the failure of running containers. This approach to orchestration results in
a system that is always trying to converge to what was specified by the user, resulting in good fault-tolerance (to the point of hiding
errors) and a clear path to implement continuous integration.

Other essential features are the ability to create namespaces (that we use to deploy several versions concurrently),
name resolution based on \texttt{DNS}, and a flexible approach to storage provisioning.

The main drawback of \texttt{kubernetes} is that it draws heavily on concepts and terminology of systems engineering. Its low
abstraction level can potentially be an obstacle for end user developers. However, since \texttt{kubernetes} is offered by many 
\emph{cloud} \cite{fehling2014cloud} providers, and also supports batch workloads, it could be a good alternative for researchers who have limited access
to high performance computing resources. Instead of creating clusters based on \emph{virtual machines}, it could
be more cost effective to run multiple containers~\cite{Julian:2016:CRI:2949550.2949562}.

\subsection{DevOps and CI}
\label{sec:devops}
DevOps   (a   portmanteau   of   ``development''   and ``operations'')  is  a  software  development  method  that  extends the  agile  philosophy  to  rapidly  produce  software  products  and services  and  to  improve  operations  performance  and  quality assurance~\cite{huttermann:2012:devops}.
DevOps tries to improve the integrated planning and execution of delivering new value with several guidelines: 
\begin{enumerate*}[label=(\roman*)]
    \item Make developers production-aware and vice-versa; 
    \item Frame developers and operators adopt the same toolset; 
    \item Short the release cycles so that the impact of releases is more manageable; and, 
    \item Promote tools to automate the setup of environments to make them reproducible.
\end{enumerate*}
DevOps aims at a continuous pipeline of delivery where new features are automatically tested in the correct environment and then approved for production.

In past projects we were already using some \texttt{DevOps} approaches like using standardized build and testing
environments supported by virtual machines to reproduce the client's environment with as much fidelity as possible.

However, this project was the first where we fully embraced the \emph{Infrastructure as Code} (\texttt{IaC}) ~\cite{6265084}. 
Our strategy is heavily based on the \texttt{git} distributed version control tool, both on the microservice and the integration level.

After a few iterations we settled on having three different versions of the prototype running at the same time. 
One called the \texttt{dev} environment would run the latest version the microservices with no guarantees of \texttt{API} stability. 
The second environment, \texttt{stable} would be used for integration testing and the teams were expected to maintain backward compatibility in their \texttt{APIs} or coordinate changes with their clients. 
Finally, the \texttt{prod} environment would be exposed to our research partners (the client).

This strategy was supported at the microservice level by an agreement of all developers to maintain at least three standard branches in their repositories: \texttt{master}, \texttt{stable} and \texttt{prod}. 
All these branches would also contain \texttt{Dockerfiles} to build images from them. 
At the same time, we configured a \texttt{Jenkins} \texttt{CI} server to watch these branches for changes. 
Every time a change was detected, a \texttt{Jenkins} job would check out the latest version, build the container image and submit it to a central registry. 
The naming scheme adopted was \texttt{registryserver/microservice\_name:git\_revision} to create a mapping between images and \texttt{git} revisions.

At the integration level we created a \texttt{kubernetes} configurations repository. 
To automate the task of submitting the configurations to kubernetes,
we developed a simple \texttt{Python} script around the \texttt{kubectl CLI} that would not only send the declarative \texttt{YAML} configuration files to the \texttt{kube-apiserver} but also create \texttt{configmaps} from configuration files and
ensure that the services would be restarted if the configuration changed. 
It also supports environment-specific configurations by leveraging the \texttt{Jinja2} templating library to bind configuration variables, but in general we tried to avoid differences between versions as much as possible.

\begin{figure}[!htb]
\centering
 \includegraphics[width=0.4\textwidth]{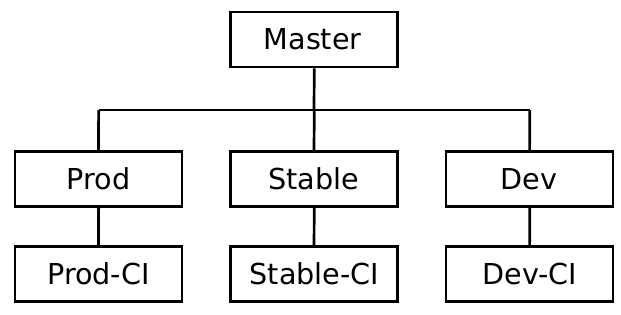}
 \caption{IaC repository}
 \label{fig:iac_repo}
\end{figure}

This orchestration repository had six branches: \texttt{dev}, \texttt{stable}, \texttt{prod}, \texttt{dev-ci}, \texttt{stable-ci}, \texttt{prod-ci}, as shown
in Figure \ref{fig:iac_repo}.
Each pair of \texttt{env} and \texttt{env-ci} branches worked in tandem to support \texttt{CI} to the three environments. 
The main branch was meant for human operators to edit, while the \texttt{*-ci} branch was maintained by a \texttt{Jenkins} job. 
In practice, every time a developer committed code, it would trigger a container build job which in turn would trigger the deployment job that would do the follow:
\begin{enumerate*}[label=(\roman*)]
    \item Take the latest revision of the \texttt{*-ci} branch;
    \item \label{item:ci-merge} Merge the latest changes from the main branch;
    \item Edit the line containing the image name of the microservice to reflect the latest revision;
    \item \label{item:ci-commit} Commit and push the changes; and, 
    \item \label{item:ci-deploy} run the deployment script.
\end{enumerate*}
Another job consisting of steps \ref{item:ci-merge}, \ref{item:ci-commit} and \ref{item:ci-deploy} could also be triggered by changes to the orchestration repository.

We found that the advantage of this scheme is that we can track all changes to the global state of the prototype because the orchestration repository
tracks the revisions of the microservices. This scheme also put the decision of publishing changes to any environment entirely in the hands of the
microservice developers. Nevertheless, this scheme is probably only applicable when all teams are part of the same organization, as is our case.

Despite the successful application of \texttt{CI} to program code, there is one significant challenge that we still have not addressed: the \texttt{CI} of data
and data \emph{schemata}. 
One of the microservices uses \texttt{RDF} \cite{RDF} to represent domain ontologies and data, which are stored in a \emph{triple store}.
Although the ontologies are modeled using \texttt{OWL} \cite{OWL} and despite the fact that these files are also stored in a git repository following the 
same branch scheme, we have not been able yet to develop a way to apply ontology changes automatically. 
In other databases in the project the problem also exists, but the data \texttt{schema} is much simpler and has been very stable over time.

\subsection{Logging, Monitoring \& Debugging}
\label{sec:monitoring}
It is generally accepted that enforcing correctness of software is more difficult in distributed systems than in single-process applications.
We argue that in microservices it can be even harder because every team is free to use the languages and tools that they deem best for the task. 
In this scenario, the usefulness of specialised debugging tools is restricted to the cases where a developer can reproduce the error in a controlled environment on her development machine, and her tools and debugging knowledge is not necessarily applicable to other microservices.

There are seven accepted techniques that can help in reducing errors in distributed systems~\cite{Beschastnikh}: testing, model checking, theorem proving, record and replay, tracing, log analysis, and visualisation. 
Apart from automated testing that some of our microservices implement in their \texttt{CI} pipelines, we employ record and replay, some tracing, log analysis, and visualisation.

The foundation of our strategy is to continuously record application logs and performance metrics on all servers. 
We accumulate these logs and metrics in a database, \texttt{elasticsearch}, and make them available for searching in the form of graphs in a web-based user-interface.
We employ a logging strategy for containers~\cite{ContainerLogging} which consists of creating a log entry for each line written to the standard output.
The downside of this strategy is that multi-line execution traces are dismembered, making it harder to read them.

We have implemented custom log parsers using \texttt{logstash} log aggregation tool to enrich the logs and allow advanced filtering based on the environment, the microservice, and the server. 
In some cases we also add additional parsers to components like \texttt{nginx}, to extract rich information, including timing data about every \texttt{HTTP} request. This data supports not only debugging but also the identification of performance problems and rich visualisations. 
Since our services all use \texttt{HTTP}, we can use these logs to replay calls.
For requests coming from the user, \texttt{nginx} also generates a unique trace identifier that is passed in subsequent requests, allowing us to group all background
service calls resulting from a single user interaction.

\todo[inline]{A participacao do SME aparece apenas aqui. Seria legal tentar inclui-lo nas outras secoes ou contar a historia de uma forma que faca um contraposto com os trabalho dos SMEs}
One difficulty has been to convince novice programmers and SMEs to be consistent in their use of logging. 
In a as-a-Service deployment good logs are especially important for \emph{post-mortem} analysis and replays, since the programmer has no direct access to the running service. 
Experienced programmers who have been exposed to libraries, such as \texttt{log4j}, recognise the benefits of such an approach and produce more useful logs. 
Ideally we would like to adopt a single logging discipline across all microservices, although from the point of view of SMEs this count as unwanted accidental complexity. 
Our qualitative survey in section \ref{sec:survey} support this view since less than half of the developers of the project used structured logging approaches, although more than half declared that logs were essential for debugging.

Testing in scientific computing is challenging because it is difficult to develop a test oracle independent of the software under development~\cite{Storer:2017:BCS}. Storer~\cite{Storer:2017:BCS} reports that weather scientists test their code on models of perfectly spherical planets for which an analytical solution is known. 
We have seen this exact testing methodology on other projects.
However, in this project, a great challenge for some microservices was the dependence on large datasets for testing that cause very long testing times. 
When integrated into an automated \texttt{CI} pipeline,
this caused the running time to be so long as to defeat the purpose of continuous deployment. 
If a developer needs to fix a small bug, say a typing error, and the tests could take over an hour to run, the natural tendency is to skip testing altogether. 
However, we have seen several developers creating unit tests for self-contained portions of code.

A testing difficulty is the reliance on other services. 
\emph{Mocking} the other services is a strategy that was not adopted in this project because, during its lifetime, there have been several cases of complete \emph{API} re-designs, which made it hard to justify the investment building and maintaining mocking objects. 
Another strategy that we could have adopted is to setup entire environments with databases initialised with reference data, but we could not find the time to do it with the fast-paced development of the project.
What some microservice developers ended up doing was running automated tests with their service pointing to the services of the \texttt{stable} environment (described in Section \ref{sec:devops}).

For one of our microservices that has faster running tests the team has implemented an automated \texttt{API} verification based on parsing the \texttt{OpenAPI} 
specification files for that microservice. 
\section{Mixed-method study}
\label{sec:study}

The goal of this study is to investigate the effects of the DevOps practices and tools on the development of a large scientific information system. 
We conducted a mixed-method study composed by a questionnaire and static source code repository analysis along with some usage metrics. 
The static analysis of source code repositories aims to provide an overview of the project implementation as well as another source of evidence using data in a triangulating fashion, because the sample size is small. 

\subsection{Qualitative survey}
\label{sec:survey}
According to the design types described by Kitchenham and Pfleeger \cite{Kitchenham:2002:PSR}, the self-administered questionnaire was a case control study, since we asked ``participants about their previous circumstances to help explain a current phenomenon``. In this particular case, we asked them about their background, their knowledge on software engineering practices and tools, focusing on \texttt{DevOps} and microservices. We elaborated the questionnaire (Appendix~\ref{sec:questionnaire}) as a web form and sent an email to the members of the project asking them to answer the questionnaire. The response rate was 18 out of 29 members of the mailing list (approximately 62\%), although some people on this list had very limited involvement in the project. The questionnaire was not sent to the \texttt{SMEs} of the client company.

Out of 18 responses, 15 people declared to have a computer science background. The other 3 had formal training in design. In the group of computer scientists, only 2 declared to specialise in software engineering although a total of 12 rated their software engineering expertise as high or above average. The other specialised in a diverse set of areas like human-computer interaction, databases, computer graphics, multimedia and computer vision. Only 6 persons reported to have worked on code bases of more than 50K \texttt{LoCs}. 8 had previous experience with agile methodologies.

When asked to rate their involvement with development on a scale of 1 to 5, 15 people chose 3 or above. These people, who we will call \emph{developers}, were not exactly the same 15
people with computer science background. Among developers, only 2 had never worked with agile methodologies. 13 developers reported to be knowledgeable in git, 10 had used branching
before and 11 used \emph{feature branches} during the project.

4 developers reported to have significant experience with cloud computing, but 8 had considerable experience in client-server programming. Only 4 developers claimed to have some degree of computer
network expertise. 

When asked if they would split up microservices into smaller ones or otherwise rearrange the microservices, the majority answered that they would not change it. 
All others declared that they were not qualified to answer it. 
Only 3 people thought that the microservices developed reflected Conway's law~\cite{conway:1968}. 4 people declared to have difficulty remembering which service was called by others. Only 4 people
thought that some APIs had poor encapsulation although a total of 9 declared that internal change in the code of called services could break their own. 6 people declared being able to draw an accurate diagram of the service logical architecture.

Only 7 developers had previous docker or virtual machine experience, and only 7 developers built a docker image of their service during the project, showing that the \texttt{CI} automation was considered effective. 
A mere 3 people declared to be able to draw an accurate infrastructure diagram, while 5 reported to have made significant changes to the \texttt{IaC} code. 
Of those, all of them rated their own software engineering experience as high or above average and had a computer science background.

In general, the amount of tested code was very low, only 8 declared to have some kind of automated testing. This reflects the general difficulty in testing scientific code
reported in the literature.

7 developers declared that their code relied heavily on calls to other services and of those, 4 considered the dependency to be strongly incidental in nature (the other 3 were neutral).
A full 9 people declared they would be able to setup everything needed to run their code, which is consistent with the number of people who did not rate the dependency of other services as being high. 
Only 3 people reported to rely heavily on asking others to make progress their work. 
2 people (from the \texttt{UI} team) declared to be blocked in their work by malfunctioning of services on a daily basis.

11 people had participated in debugging activity and valued application logs for this activity. 
9 considered debugging to be harder with microservices than monolithic code and of those 5 considered it much harder. 
7 \emph{debuggers} used logging libraries or claimed to have a consistent methodology for print statements.
9 debuggers made use of the log searching tool, and of those 2 had trouble using them or considered them insufficient. 
To effectively use the log search tool, one had to be able to apply several filters, many based on \texttt{kubernetes} and \texttt{docker} concepts.

8 developers found 3 environments easy to remember and 9 had no problem using 3 branches mapped to these environments consistently. 
Only 2 declared to have difficulty discovering which version of their code was running in each environment.

\subsection{Source code metrics}
\label{sec:source-code-metrics}
In this section we present source code metrics to provide an overview of the project implementation. 
Table~\ref{tab:code} shows the size in KLOC (thousands of lines of code) and implementation languages for each service. 
The services called \texttt{MS-*} are \texttt{RESTFul} microservices, while the one called \texttt{P-*} were based on the \texttt{Kafka} message queue. 
Services were implemented using different programming languages, but python is the most commonly used (12 out of 17 services were either partially or completely implemented in python). 
The usage of python was not enforced by any technical constraint, so the reason it was chosen for implementation of many services was because many software engineers are proficient in this language. 
11 out of 18 (61\%) of the survey participants (see Section~\ref{sec:survey}) are proficient in python and some of them were involved in the development of two or more services. 
Furthermore, using the same language share information and even code among software engineers, because many services have similar quality attributes (e.g., logging, handling communication errors, Open API implementation).

\begin{table}[!htb]
 \centering
 \caption{Service's size and language}
 \label{tab:code}
 \begin{tabular}{c|c|c} \hline
 \textbf{Service}          & \textbf{KLOC} & \textbf{Language} \\ \hline
 UI               & 52.3 & Typescript, CSS, HTML\\ \hline
MS-1                & 39.3 & python               \\ \hline
MS-2               & 40.1 & python                \\ \hline  
MS-3               & 16.4 & python                \\ \hline                  
MS-4              & 11.9 & python                \\ \hline
MS-5              & 6.1  & python, R            \\ \hline  
MS-6             & 0.8& python               \\ \hline                  
MS-7  & 3.2  & python, Java             \\ \hline
MS-8 &  8.8 & Javascript            \\ \hline  
P-1 & 0.8& python                \\ \hline                  
P-2 & 0.5 & Java                 \\ \hline
P-3 & 0.6  &  Java          \\ \hline  
P-4  & 2.5  & python, Java          \\ \hline  
P-5 & 0.3& python                \\ \hline
P-6 & 3.5 & python                \\ \hline                                  
P-7  & 5.6 & python, lisp, prolog  \\ \hline                                  
 IaC               &  7.1  & yaml, other config lang. \\ \hline
\end{tabular}
\end{table}

Microservices are language agnostic because services communicate with each other via lightweight mechanisms~\cite{newman2015building},\texttt{REST APIs} in the case of this project. 
Thus, developers can use the best suited programming language for a particular task. 
For instance, the MS-5, which is heavily based on statistics, has its core implemented in \texttt{R} and the P-7 service was implemented using PROLOG and LISP, 
which are common in the artificial intelligence domain. 
The last line accounts for all \texttt{IaC} (infrastructure as a code) implementation, including \texttt{kubernetes} configuration, \texttt{Dockerfiles} and service configuration files. Notably it represents a mere 3\% of the overall \texttt{LoCs}, although the \texttt{CI} configuration in the \texttt{Jenkins} is not included.

The size of the service can vary two orders of magnitude, from a few hundreds LOC to dozens of thousands LOC, even when they are implemented using the same language. Some services are inherently more complex than others due to the large number of features provided. 

\subsection{Version control metrics}
\label{sec:vcm}
In order to assess to what extent CI tools have influenced on software engineers, we measured the impact of introducing Jenkins to their daily work routine. 
We analysed the Git logs of services implementation that started before Jenkins was made available. 
Jenkins availability started on the 1\textsuperscript{st} of June 2016 and the Git logs that were analysed since the first commit on 29\textsuperscript{th} December 2015 until 5\textsuperscript{th} of September 2018. 
Based on these Git logs, we selected the commits of 10 software engineers, which were the only that have worked on this project before and after Jenkins availability. We counted the total number of commits per week of these selected software engineers. 
Then, we divided this total commits per week by the number of software engineers that have committed at least once in the week in order to select only those that were actively working on the project and to remove those on vacation and those that left the project. 

\begin{figure}[!htb]
\centering
 \includegraphics[width=0.5\textwidth]{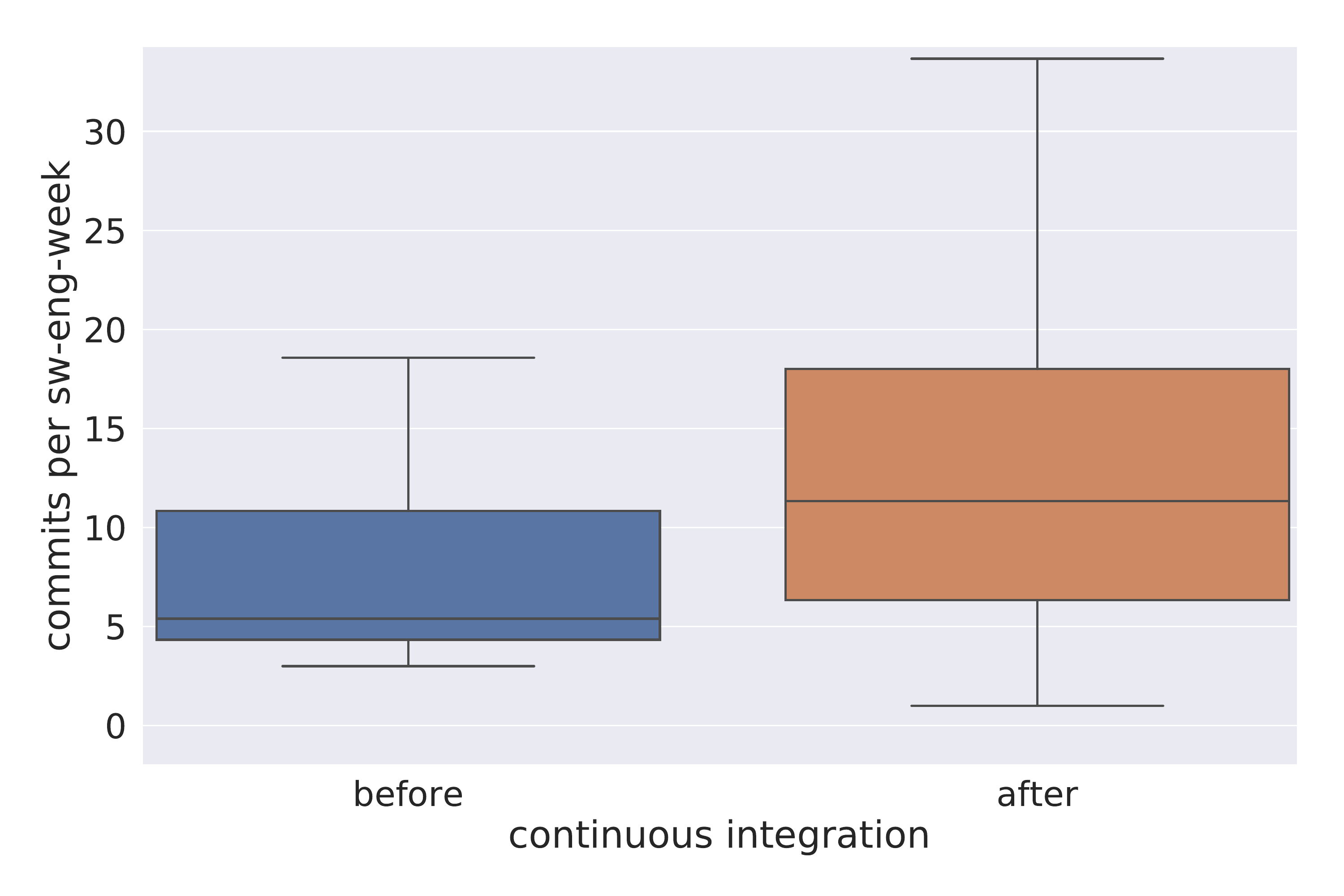}
 \caption{Average commits per software engineer-week}
 \label{fig:commits_per_week}
\end{figure}

Figure~\ref{fig:commits_per_week} shows a boxplot of the average number of commits per week of the selected software engineers. 
The average number of commits increased from a 5.4 to 11.3 median after Jenkins which introduction facilitated the deployment of the services on multiple environments. Then, when software engineers finished the implementation of new features or bug fixes they could easily deploy on a development environment to perform integration testing.  

\begin{figure}[!htb]
\centering
 \includegraphics[width=0.5\textwidth]{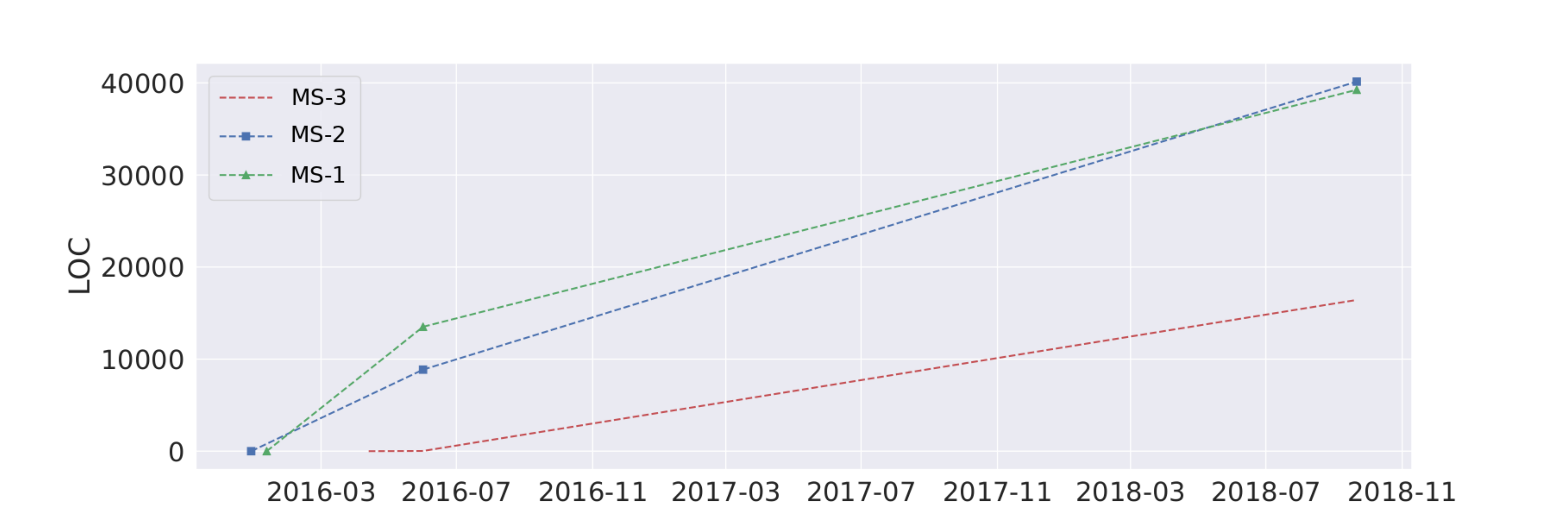}
 \caption{LOC evolution per service before and after the availability of CI tool on June 2016}
 \label{fig:LOC-evolution}
\end{figure}

Three services implementation started before Jenkins availability, namely MS-1, MS-2, and MS-3. Figure~\ref{fig:LOC-evolution} shows that the development of two of them was more intense, in terms of lines of code (LOC), before Jenkins was available than after it. Thus, it was not the development intensity that influenced the number of commits. These results are similar to the perceived benefits of DevOps tools observed in other studies, such as Itkonen \textit{et al.}~\cite{Itkonen:2016:PBA}.

\subsection{Runtime metrics}

The prototype had a total of 16 \texttt{SME} users who would test the prototype. Every week day, on average 3 of them accessed the prototype. Over a period of 3 months, we measured
a total of over 3 million \texttt{HTTP} requests, of which 72\% were completed in under 10ms (as measured by the \texttt{nginx} reverse proxy, not including user network latency).
In the same time period, we recorded over 5000 background task execution, the longest of which took almost 7 days, although only 10 took more than 24 hours. 

\subsection{Limitations of study}

The most serious threat to validity in this work is that the authors were also developers of the project and also answered the questionnaire. Since the authors account to about 20\% of the respondents, we thought that excluding their perspective would impact the validity even more by reducing the sample size and also by excluding a large part of the people with software engineering background. 

Another threat to validity in our conclusions is that our microservices evolved together to meet the requirements of a large project, instead of evolving independently as is the norm in the microservices literature. In some cases a same database is accessed by more than one microservice.


\section{Related Work}
\label{sec:related}

Idaszak \textit{et al.}~\cite{Idaszak:2017:Hydroshare} describe the lessons learned when applying \texttt{Dev\-Ops}, among other techniques, to a large scientific software project. 
In this project, a software engineer was appointed as \texttt{DevOps} lead, which allowed SMEs to learn how to use tools like Git and Github, thus simplifying the contributions and increasing collaboration. They concluded that software engineering techniques will foster research by ensuring validity of findings in a long run.

Itkonen \textit{et al.}~\cite{Itkonen:2016:PBA} conducted interviews with both the supplier organization and their customer to understand how the customer perceived the changes to process and methods, and to identify how the adoption of continuous delivery has had beneficial effects on their work. Their findings were the following: (i) automated testing reduced bugs in production; (ii)~development and production environments are similar, which reduced the risk of errors; (iii)~simple metrics, such as lead times and release cycle metrics do not give good enough picture of how development works.

Savor \textit{et al.}~\cite{savor2016continuous} analysed data from Git repositories from Facebook and OANDA companies over a period of 7 and 5 years, respectively, in order to investigate the experiences with \texttt{CI} in these companies. Their findings were that \texttt{CI} allows scaling the number of developers and code base size and also the developers prefer short release cycles over long ones.

Subjectively comparing our project to the ones described in this section and also to the ones observed by Segal~\cite{4351335} and Carver \textit{et al.}~\cite{4222616} and the surveys by Johanson and Hasselbring~\cite{Johanson} and Storer~\cite{Storer:2017:BCS} we believe that it is representative of scientific computing in general, although there are three important disclaimers to be made: (i)~since we do not depend on research grants, our incentive structures are slightly different than in academia in general; (ii)~our team does have a large percentage of people with a computer science background; and (iii)~compared to the literature we have the advantage of time, since \texttt{Agile} and \texttt{\texttt{DevOps}} already have broad acceptance in the industry and are already past the "peak of inflated expectations"~\cite{hype}.
\section{Conclusion}
\label{sec:conclusion}

In scientific information systems development, scientists (SME) and software engineers should work close together to create the multidisciplinary scientific code having high quality of delivery. 
We presented our practical experience towards bridging the gap in scientific information systems development concerning the use of modern software engineering in scientific computing.
We explored the development of a scientific system applying microservices, \texttt{DevOps}, and as-a-service approach delivery from the start to a large applied research project involving a multidisciplinary team of more than 30 people over 3 years. 

Our main goal was to validate the hypothesis that the use of microservices and DevOps would allow computational scientists to work in the more minimalistic prototype-oriented way that they prefer while the software engineering team would handle the integration. 

The project was considered successful and has been awarded innovation prizes in both ours and the client's Oil\&Gas company.
We demonstrated that the adoption of \texttt{CI} and as-a-service delivery has positively contributed to this, based
on previous industry projects where we did not apply these techniques. 
In these other projects, the difficulty of delivering new developments reduced the rate at which client SMEs could test and validate the software, thereby reducing the speed of iterations.

Internally, the \texttt{CI} pipeline also made the overall process easier since people were relieved of the burden of deploying their software manually. 
In some previous projects, different teams would manually install their software by hand on a shared machine for integration. 
There was no tracking of what was installed and what were the steps to install it, making any update a risky activity because one could inadvertently break things and not know how to revert to a ``sane'' state. 
In contrast, now that almost anything is tracked and reversible, the barrier to test new commits is much lower.

The survey that we conducted also showed that only a small part of the overall team needs to have expert knowledge of \texttt{CI} and \texttt{DevOps} for these methodologies to work. 

\bibliographystyle{unsrt}  
\bibliography{bibliography, researchops} 

\begin{thebibliography}{10}

\bibitem{4351335}
J.~Segal.
\newblock Some problems of professional end user developers.
\newblock In {\em IEEE Symp. on Visual Languages and Human-Centric Computing
  (VL/HCC 2007)}, pages 111--118, Sept 2007.

\bibitem{Johanson}
A.~Johanson and W.~Hasselbring.
\newblock Software engineering for computational science: Past, present,
  future.
\newblock {\em Comput. in Science Engineering}, 20(2):90--109, 2018.

\bibitem{Storer:2017:BCS}
Tim Storer.
\newblock Bridging the chasm: A survey of software engineering practice in
  scientific programming.
\newblock {\em ACM Comput. Surv.}, 50(4), 2017.

\bibitem{anglia}
House of~Commons~Science and Technology~Committee 2010.
\newblock Eighth report of session 2009-10.

\bibitem{4222616}
J.~C. Carver, R.~P. Kendall, S.~E. Squires, and D.~E. Post.
\newblock Software development environments for scientific and engineering
  software: A series of case studies.
\newblock In {\em 29th International Conference on Software Engineering
  (ICSE'07)}, pages 550--559, May 2007.

\bibitem{wiki:Complexity}
Wikipedia.
\newblock {Complexity} --- {W}ikipedia{,} the free encyclopedia.
\newblock
  \url{http://en.wikipedia.org/w/index.php?title=Complexity&oldid=856416658},
  2018.
\newblock [Online; accessed 21-September-2018].

\bibitem{1702388}
T.~J. McCabe.
\newblock A complexity measure.
\newblock {\em IEEE Transactions on Software Engineering}, SE-2(4):308--320,
  1976.

\bibitem{halstead}
Maurice~Howard Halstead.
\newblock Elements of software science, 1977.

\bibitem{Ackroyd}
Karen~S. Ackroyd, Steve~H. Kinder, Geoff~R. Mant, Mike~C. Miller, Christine~A.
  Ramsdale, and Paul~C. Stephenson.
\newblock Scientific software development at a research facility.
\newblock {\em IEEE Software}, 25(4):44--51, 2008.

\bibitem{Kleb}
William~A. Wood and William~L. Kleb.
\newblock Exploring xp for scientific research.
\newblock {\em IEEE Softw.}, 20(3):30--36, 2003.

\bibitem{savor2016continuous}
Tony Savor, Mitchell Douglas, Michael Gentili, Laurie Williams, Kent Beck, and
  Michael Stumm.
\newblock Continuous deployment at facebook and oanda.
\newblock In {\em Proceedings of the 38th International Conference on Software
  Engineering Companion}, ICSE '16, pages 21--30, New York, NY, USA, 2016. ACM.

\bibitem{gimenez2020devops}
Paulo J.~A. Gimenez and Gleison Santos.
\newblock Devops maturity diagnosis--a case study in two public organizations.
\newblock In {\em XVI Brazilian Symposium on Information Systems (SBSI 2020)},
  pages 1--8, 2020.

\bibitem{fowler2014microservices}
Martin Fowler and James Lewis.
\newblock Microservices.
\newblock https://martinfowler.com/articles/microservices.html, 2014.
\newblock Accessed on: Feb-2021.

\bibitem{huttermann:2012:devops}
Michael H{\"u}ttermann.
\newblock {\em DevOps for Developers}, volume~1.
\newblock Springer, 2012.

\bibitem{newman2015building}
Sam Newman.
\newblock {\em Building microservices}.
\newblock O'Reilly Media, Inc., 2015.

\bibitem{zimmermann2016microservices}
Olaf Zimmermann.
\newblock Microservices tenets.
\newblock {\em Computer Science-Research and Development}, pages 1--10, 2016.

\bibitem{5337641}
S.~Killcoyne and J.~Boyle.
\newblock Managing chaos: Lessons learned developing software in the life
  sciences.
\newblock {\em Computing in Science Engineering}, 11(6):20--29, 2009.

\bibitem{MongoDB}
B.~Bradshaw and C.~Chodorow.
\newblock {\em MongoDB: The Definitive Guide, 3rd Edition}.
\newblock O'Reilly Media, Inc., 2018.

\bibitem{stricker2018microservices}
Robert Stricker, Daniel M{\"u}ssig, and J{\"o}rg L{\"a}ssig.
\newblock Microservices for redevelopment of enterprise information systems and
  business processes optimization.
\newblock In {\em Intenational Conference on Enterprise Information Systems
  (ICEIS)}, pages 719--726, 2018.

\bibitem{Fielding:2000:ASD}
Roy~Thomas Fielding.
\newblock {\em Architectural styles and the design of network-based software
  architectures}.
\newblock PhD thesis, Univ. of California, 2000.

\bibitem{Burns:2018:DDS}
Brendan Burns.
\newblock {\em Designing Distributed Systems: Patterns and Paradigms for
  Scalable, Reliable Services}.
\newblock O'Reilly Media, Inc., 1st edition, 2018.

\bibitem{JWT}
Sebasti\'{a}n Peyrott.
\newblock Jwt handbook.

\bibitem{datadog}
Datadog.
\newblock 8 surprising facts about real docker adoption.
\newblock https://www.datadoghq.com/docker-adoption/, 2018.
\newblock Accessed on: Fev-2021.

\bibitem{DOCKER}
Dirk Merkel.
\newblock Docker: lightweight linux containers for consistent development and
  deployment.
\newblock {\em Linux Journal}, 2014(239):2, 2014.

\bibitem{6265084}
D.~Spinellis.
\newblock Don't install software by hand.
\newblock {\em IEEE Software}, 29(4):86--87, 2012.

\bibitem{Boettiger:2015:IDR:2723872.2723882}
Carl Boettiger.
\newblock An introduction to docker for reproducible research.
\newblock {\em SIGOPS Oper. Syst. Rev.}, 49(1):71--79, January 2015.

\bibitem{kubernetes}
Brendan Burns, Brian Grant, David Oppenheimer, Eric Brewer, and John Wilkes.
\newblock Borg, omega, and kubernetes.
\newblock {\em Queue}, 14(1):10:70--10:93, January 2016.

\bibitem{kubernetes2}
Kelsey Hightower, Brendan Burns, and Joe Beda.
\newblock {\em Kubernetes: Up and Running Dive into the Future of
  Infrastructure}.
\newblock O'Reilly Media, Inc., 1st edition, 2017.

\bibitem{fehling2014cloud}
Christoph Fehling, Frank Leymann, Ralph Retter, Walter Schupeck, and Peter
  Arbitter.
\newblock {\em Cloud computing patterns: fundamentals to design, build, and
  manage cloud applications}.
\newblock Springer Science \& Business Media, 2014.

\bibitem{Julian:2016:CRI:2949550.2949562}
Spencer Julian, Michael Shuey, and Seth Cook.
\newblock Containers in research: Initial experiences with lightweight
  infrastructure.
\newblock In {\em Proceedings of the XSEDE16 Conference on Diversity, Big Data,
  and Science at Scale}, XSEDE16, pages 25:1--25:6, New York, NY, USA, 2016.
  ACM.

\bibitem{RDF}
World Wide Web~Consortium (W3C).
\newblock Resource description framework (rdf) model and syntax specification.

\bibitem{OWL}
World Wide Web~Consortium (W3C).
\newblock Owl 2 web ontology language document overview (second edition).

\bibitem{Beschastnikh}
Ivan Beschastnikh, Patty Wang, Yuriy Brun, and Michael~D. Ernst.
\newblock Debugging distributed systems.
\newblock {\em Commun. ACM}, 59(8):32--37, 2016.

\bibitem{ContainerLogging}
Andre Newman.
\newblock Top 5 docker logging methods to fit your container deployment
  strategy.

\bibitem{Kitchenham:2002:PSR}
Barbara~A. Kitchenham and Shari~Lawrence Pfleeger.
\newblock Principles of survey research part 2: Designing a survey.
\newblock {\em SIGSOFT Softw. Eng. Notes}, 27(1):18--20, 2002.

\bibitem{conway:1968}
Melvin~E. Conway.
\newblock How do committees invent?
\newblock {\em Datamation}, 1968.

\bibitem{Itkonen:2016:PBA}
Juha Itkonen, Raoul Udd, Casper Lassenius, and Timo Lehtonen.
\newblock Perceived benefits of adopting continuous delivery practices.
\newblock In {\em Proceedings of the ACM/IEEE International Symposium on
  Empirical Software Engineering and Measurement}, ESEM '16, pages 42:1--42:6,
  New York, NY, USA, 2016. ACM.

\bibitem{Idaszak:2017:Hydroshare}
Ray Idaszak, David~G. Tarboton, and Hong~Yi {\textit{et al.}}
\newblock Hydroshare - a case study of the application of modern software
  engineering to a large distributed federally-funded scientific software
  development project.
\newblock In {\em Software Engineering for Science}. Taylor and Francis, 2017.

\bibitem{hype}
Inc. Gartner.
\newblock Gartner hype cylce.

\end{thebibliography}

\appendix

\section{Questionnaire}
\label{sec:questionnaire}

\begin{enumerate}
    \item Do you have, or are you pursuing a degree in Computer Science, Computer Engineering, Information Systems or any area closely related to software engineering?
        \begin{enumerate}
            \item If not, what is the area of your degree?

            \item If yes, do you have or are you pursuing a MSc or PhD degree?

            \item. If yes, what is the area of your Msc/PhD degree

        \end{enumerate}
    \item What was the level of experience you had with source control systems, such as git, before the project?
    \item What was the level of experience you had with cloud computing before the project?
    \item What was the level of experience you had with client-server applications before the project?
    \item Did you ever interact with deployment (docker/kubernetes/jenkins) code or configurations?
    \item What is your level of satisfaction with these tools?
    \item How much of your working time was dedicated to this project?
    \item How much of your code has automated tests?
    \item Did other services have automated tests?
    \item How often do you include logs or prints in your code?
    \item Have you worked in the development of software with multiple version branches and deployment environments before this project?
    \item Did you find it easy to maintain the master/stable/prod branches and environments?
    \item How much did you use feature branches?
    \item How frequently did you check the project's common issues?
    \item How frequently did you use git hub to track feature issues?
\end{enumerate}

\end{document}